\documentclass[a4paper,11pt]{article}
\pdfoutput=1
\usepackage[sort&compress,numbers]{natbib}

\usepackage[colorlinks=true,urlcolor=blue,
anchorcolor=blue,citecolor=blue,filecolor=blue,
linkcolor=blue,menucolor=blue,
linktocpage=true,pdfa=true]{hyperref}

\usepackage{graphicx, amssymb, amsmath}
\graphicspath{{figs/}}
\newcommand{\figw}[2]{\includegraphics[width=#1\textwidth]{#2}}

\usepackage[left=2cm, right=2cm, top=1.9cm, bottom=0.75cm,
footskip=48pt, headsep=20pt, includehead, includefoot]{geometry}
\begin{document}
\title{A Neural Network Approach for Selecting Track-like Events in Fluorescence
	Telescope Data}
\author{Mikhail~Zotov$^{1*}$, Denis~Sokolinskii$^2$
	on behalf of the JEM-EUSO collaboration\\
	$^{1*}$Skobeltsyn Institute of Nuclear Physics,\\
	Lomonosov Moscow State University,\\
	Moscow 119991, Russian Federation\\
	$^2$Faculty of Physics,
	Lomonosov Moscow State University,\\
	Moscow 119991, Russian Federation\\[5mm]
	$^*$\texttt{zotov@eas.sinp.msu.ru}
	}
\date{}
\maketitle

\begin{abstract}

	In 2016--2017, TUS, the world's first experiment for testing the
	possibility of registering ultra-high energy cosmic rays (UHECRs) by
	their fluorescent radiation in the night atmosphere of Earth was
	carried out. Since 2019, the Russian-Italian fluorescence telescope
	(FT) Mini-EUSO (``UV Atmosphere'') has been operating on the ISS. The
	stratospheric experiment EUSO-SPB2, which will employ an FT for
	registering UHECRs, is planned for 2023.  We show how a simple
	convolutional neural network can be effectively used to find
	track-like events in the variety of data obtained with such
	instruments.

\end{abstract}
\section{Introduction}

In recent years, neural networks of various configurations have been
increasingly used to analyze data obtained with fluorescent and
Cherenkov telescopes. In particular, a whole series of studies dedicated
to the analysis of gamma-ray astronomy data with neural networks has
been performed by the VERITAS~\cite{Flanagan:202199},
TAIGA~\cite{Polyakov:20214V, 2022dlcp.confE..16P}, and
CTA~\cite{Aschersleben:2021nj, Abe:2021Yg} collaborations.  Typical
tasks are the recognition of particular signal patterns in the data
flow.  In the simplest case, the problem can be reduced to classifying
data into two groups: data samples that contain a signal of the desired
type and all the rest.

Since data obtained with the help of telescopes can naturally be
considered as images or animations, one of the popular tools for
classifying them are convolutional neural networks (CNNs), created
primarily for image classification.  CNNs have demonstrated the highest
efficiency in this class of problems, see, for example,~\cite{Lecun1998,
2012arXiv1202.2745C}.  Previously, we used a CNN to recognize two types
of events recorded during the TUS experiment, the first orbital mission
aimed at testing the possibility of detecting ultra-high energy cosmic
rays (UHECRs, energies above 50~EeV) from fluorescent and Cherenkov
radiation that occurs in the Earth’s atmosphere during the development
of extensive air showers (EASs)~\cite{SSR2017, IzvRan2017, UHECR2016,
JCAP2017}.  It turned out that even the simplest CNNs with only one
convolutional layer are able to solve this problem
effectively~\cite{BMY2020, universe7070221}.  Moreover, a CNN trained
with samples prepared using conventional signal selection algorithms,
was able to detect weak signals examples of which were not included in
the training data.  This significantly expanded our understanding of
signals of the studied types and provided answers to some open
questions.

When studying UHECRs using fluorescent telescopes, the main object of
the search are track-like events in which a signal with a specific light
curve propagates along a quasi-linear track.  It is straightforward to
train a convolutional neural network to recognize such signals on
simulated data. However, the TUS data analysis has revealed that the
result of application of such a CNN to real experimental data can be
unsatisfactory if the absolute calibration of the instrument is known
with a considerable error and the background illumination is strongly
variable.

Fortunately, there is a fairly large set of track-like events in the
data of the Russian-Italian orbital experiment Mini-EUSO (``UV
Atmosphere''), which has been operating on the ISS since 2019.  These
signals have a completely different nature than signals from EASs but
some observational properties of them are the same.  We mean the glow of
meteors burning up in the atmosphere. This type of signals develops on a
much longer time scale. However, the emission curves resemble a Gaussian
shape and the signal itself propagates quasi-linearly in the field of
view of the telescope, similar to the case of EASs. In this paper, we
briefly report results of the development of a CNN for recognizing
signals from meteors in the data of Mini-EUSO.
Results of an application of the CNN to the data analysis will be
reported elsewhere.

\section{Mini-EUSO Experiment}

One of the main scientific objectives of the Mini-EUSO experiment is to
obtain a map of the glow of the Earth's night atmosphere in the near
ultraviolet wavelength band (290--430~nm), which is necessary for the
development of the next-generation orbital experiments
K-EUSO~\cite{KEUSO-2022Univ} and POEMMA~\cite{POEMMA:2020ykm}.  A rich
variety of signals in the UV range are recorded in the course of the
experiment along with the background radiation of the atmosphere, among
them anthropogenic illumination, transient atmospheric phenomena,
meteors, EAS-like signals, and many others, see,
e.g.,~\cite{2021ApJS..253...36B, Casolino:2021jO}.

The main elements of the telescope are two Fresnel lenses 25~cm in
diameter and a focal surface (FS) consisting of 36 Hamamatsu R11265-M64
multi-anode PMTs (MAPMTs) arranged in a $6\times6$ matrix. Each MAPMT
consists of $8\times8$ pixels with a single pixel size of
$2.88~\text{mm}\times2.88~\text{mm}$. The field of view of Mini-EUSO is
44~degrees, so that each pixel observes an area of approximately
$6.3~\text{km}\times6.3~\text{km}$ from the orbit of the ISS.  The time
resolution equals 2.5~$\mu$s.  The data collection is carried out
simultaneously in three modes: with the minimum time resolution and the
length of each record equal to 128 time steps, in the mode with signal
integration over 128 cycles, and also in the mode with integration over
$128\times128$ cycles, i.e., with a time step of 40.96~ms.  In the
latter case, the recording is continuous, without a trigger. It is in
this mode that it becomes possible to record full meteor tracks. We
operate only with such data in what follows.

Observations are performed approximately twice a month. In this article,
we use data obtained during eight sessions held between November~19,
2019, and April~1, 2020.

\section{Neural Network}

\begin{figure}[!ht]
	\centering
	\figw{.8}{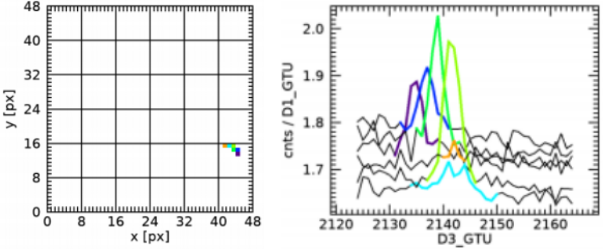}

	\caption{An example of a meteor signal registered by Mini-EUSO
	(from~\cite{Casolino:2021jO}). The left panel shows the trace of the
	meteor on the FS, with other simultaneous signals omitted. The right
	panel shows the shape of signals in active pixels.}

	\label{example}
\end{figure}

Two meteor datasets were obtained within the JEM-EUSO collaboration using the
sessions considered in the present paper. These datasets together with results
of an independent analysis performed by the authors provided a sample of
approximately 1100 events that were used as the training set.
Figure~\ref{example} from~\cite{Casolino:2021jO} provides an example of a bright
meteor signal recorded by Mini-EUSO. The left panel shows the location of active
pixels on the FS, the right panel demonstrates the respective signals.  Notice
that the signal is concentrated in only six of the total 2304 pixels while
strong and spatially extended illumination can be present in other parts of the
focal surface.

A CNN that worked well for selecting two types of signals in the TUS
telescope data consisted of one convolutional layer, one pooling layer,
and two fully connected layers~\cite{universe7070221}.  Data in input
samples had a dimension of $M\times M\times N$, where~$M\times M$
corresponded to the size of the FS, and~$N$ was the number of
``snapshots'' of the FS taken for the analysis. We tried the same
approach in the present study with $M\times M=48\times48$, which is
equal to the size of images taken by the instrument, and~$N=8,\dots,64$,
which corresponds to the range of meteor signal durations.  The data was
augmented by image rotation to increase the number of meteor signals in
the training sample.  It turned out however that this approach does not
provide satisfactory results regardless of hyperparameters of the CNN
and values of~$N$.  While training accuracy could reach 0.999, area
under the ROC-curve (AUC) calculated for testing samples hardly
reached~0.6, which is only slightly better than the result of a random
choice.  (Recall that AUC is interpreted as the probability that the NN
will make the right choice if it is given two randomly selected examples
of both classes.)

\begin{figure}[!ht]
	\centering
	\figw{.47}{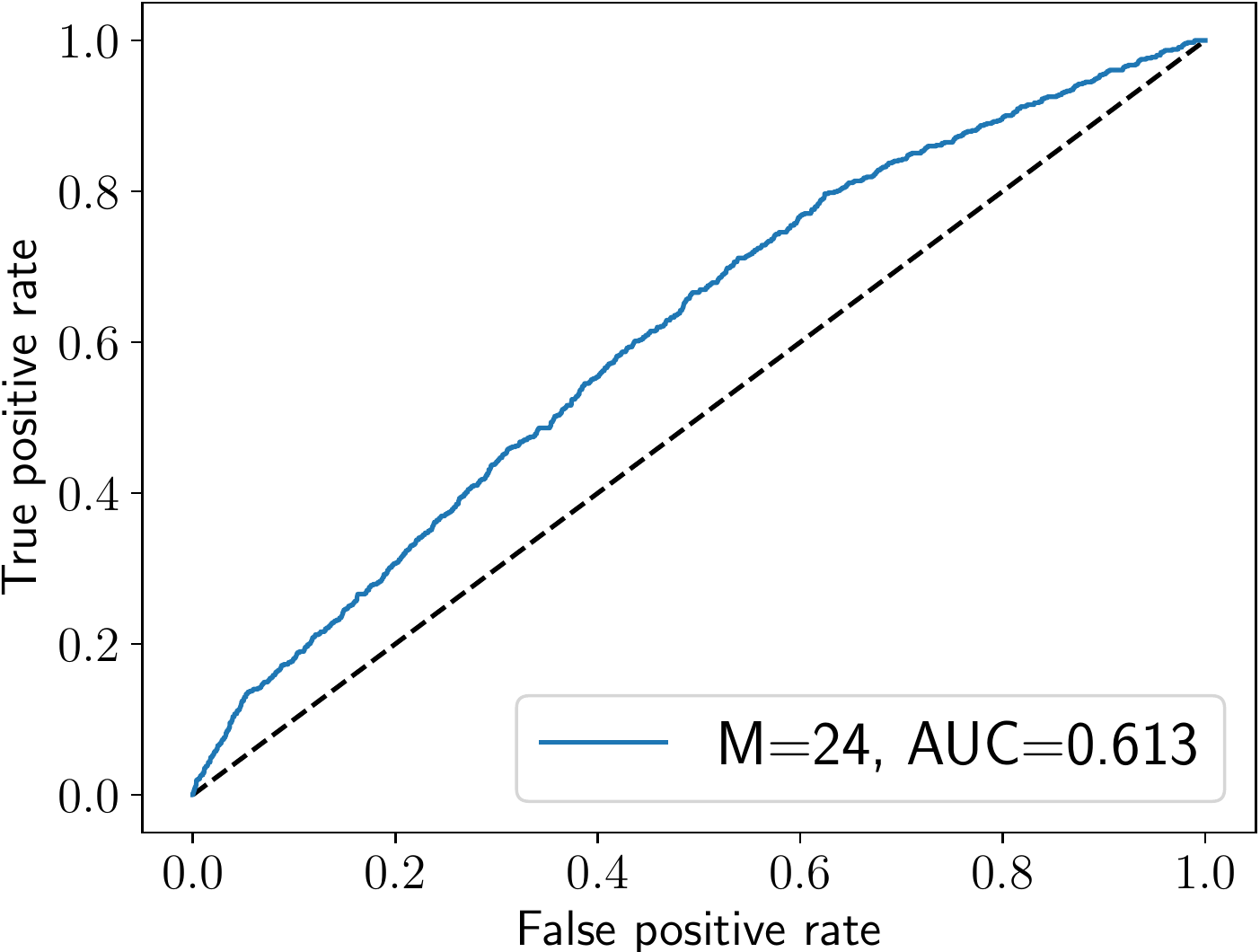}\quad\figw{.47}{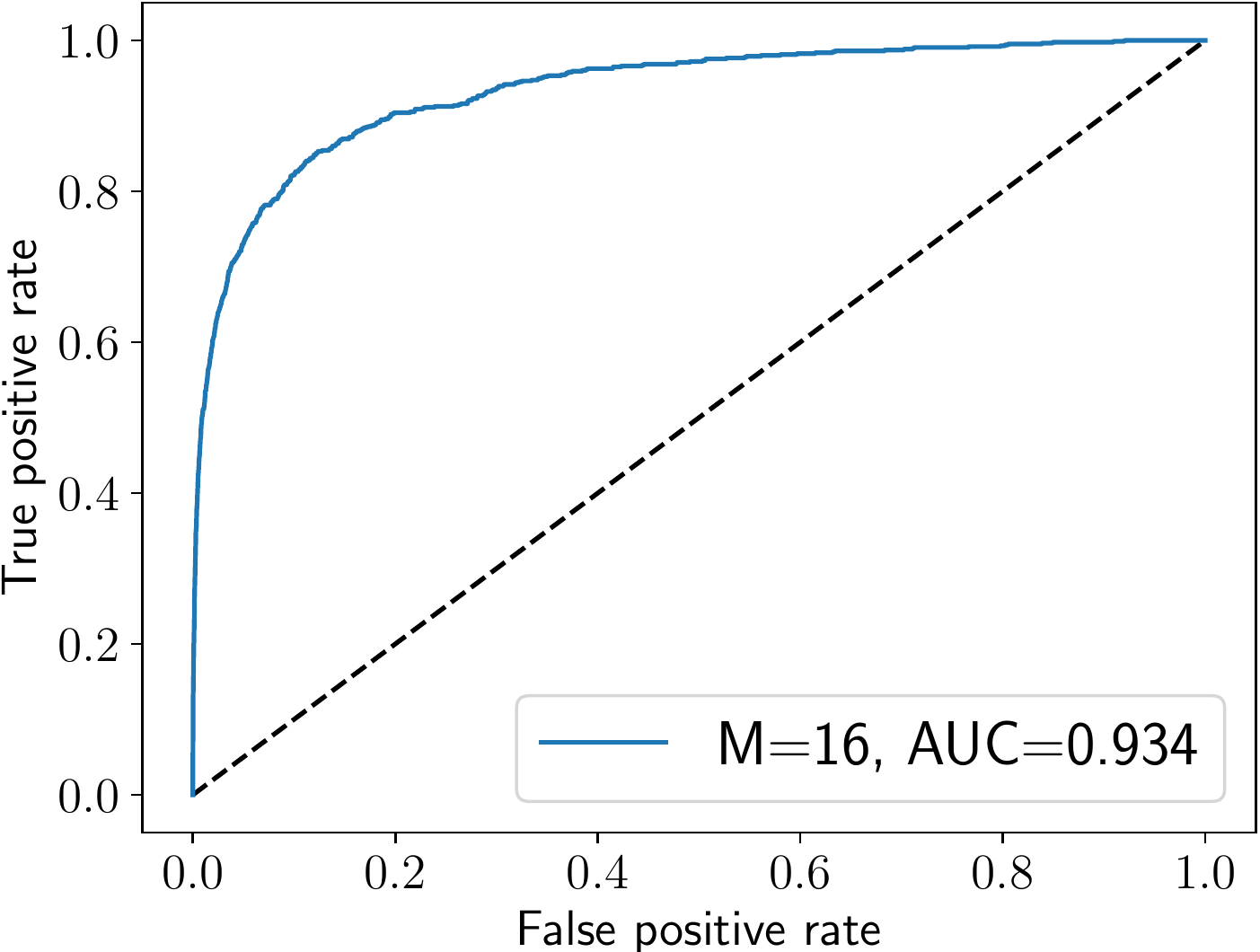}

	\bigskip

	\figw{.47}{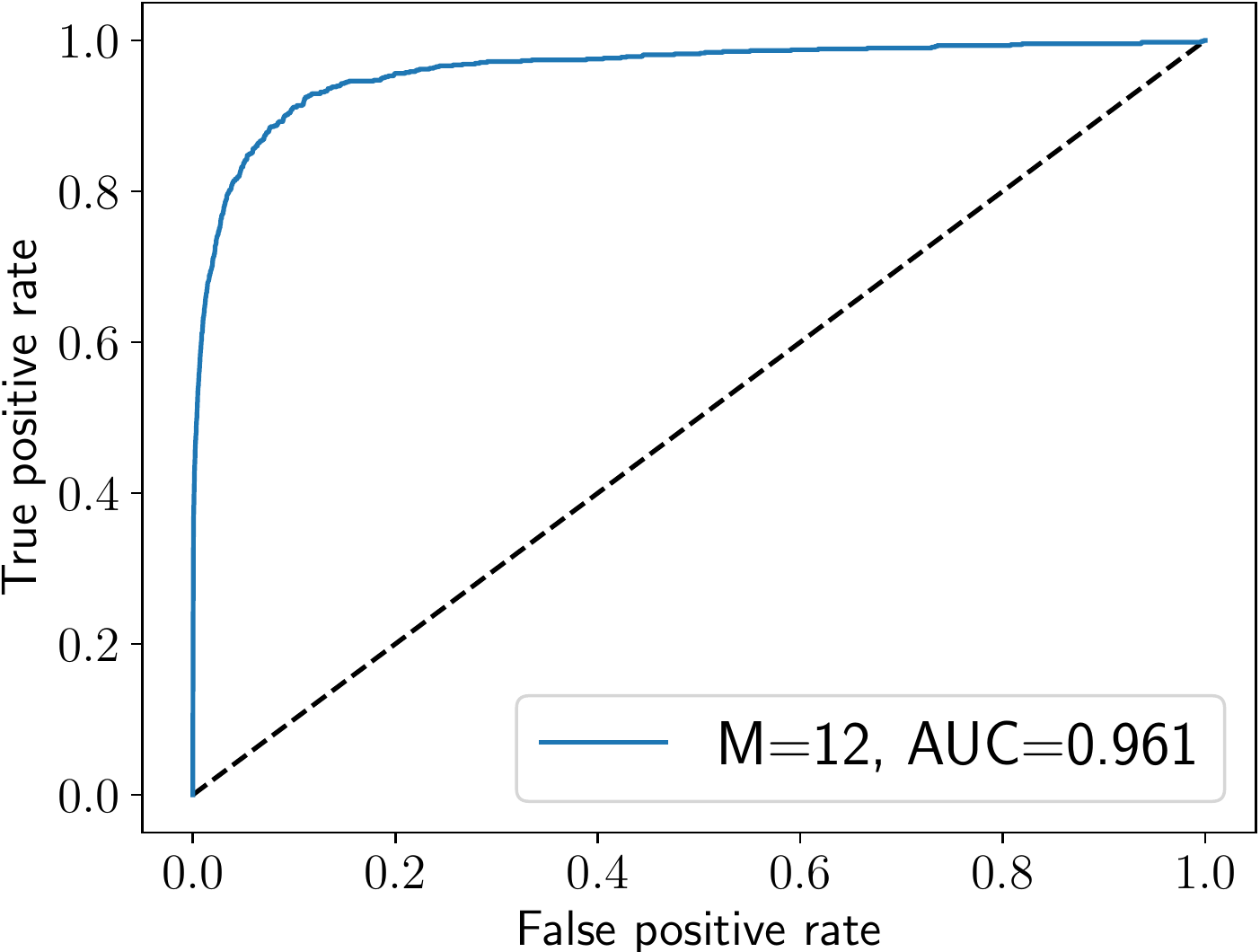}\quad\figw{.47}{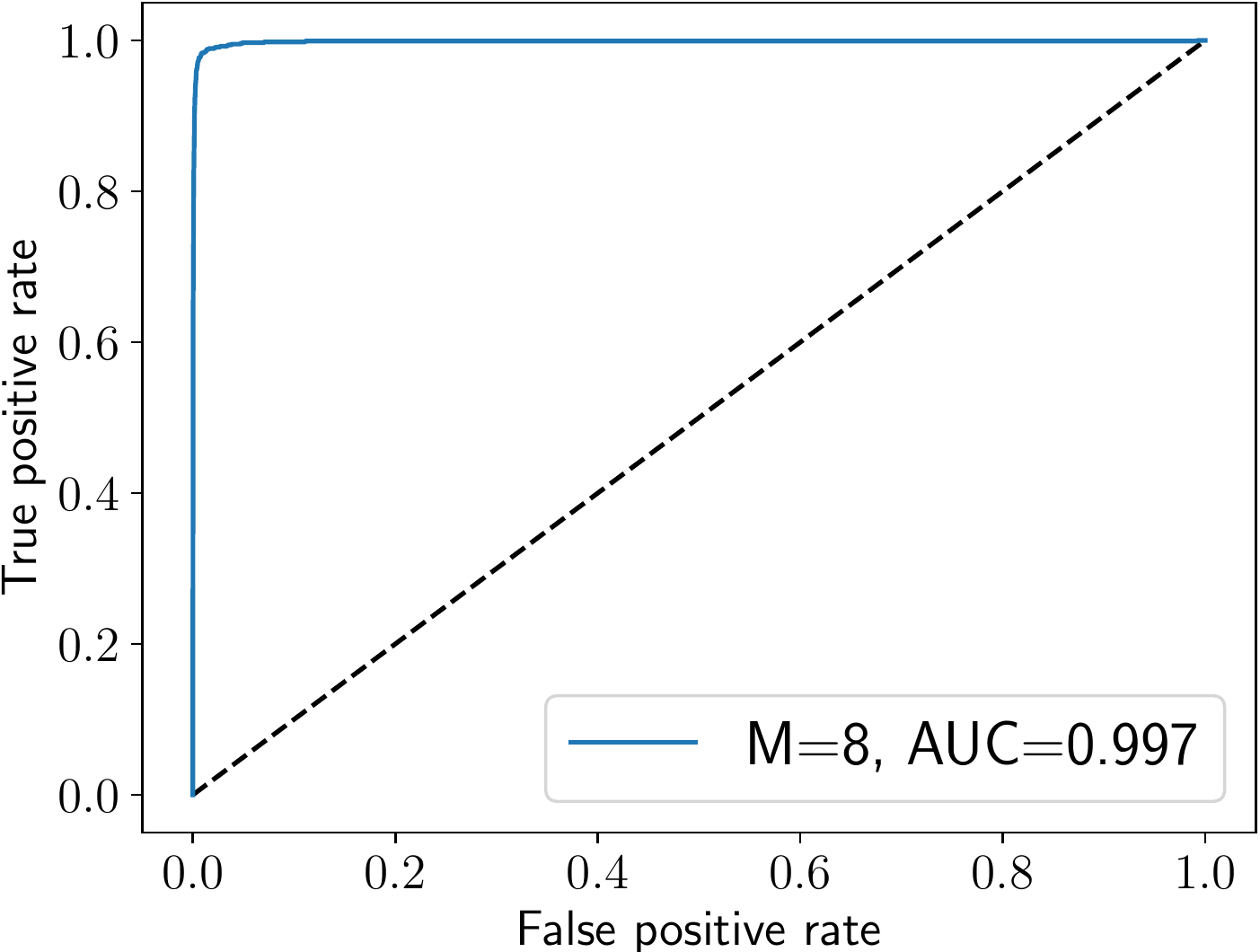}

	\caption{Dependence of ROC curves (solid lines) and AUC calculated
	for testing samples on the value of parameter~$M$, see the text for
	details.  The dashed lines correspond to the case of a random choice
	with equal probabilities.}

	\label{MxM}
\end{figure}

We tested more complex architectures of CNNs as well as LSTM networks but failed
to achieve AUC above~0.75. This led us to the idea to try smaller values $M=24,
16, 12, 8$. Input samples were created by sliding a window of the respective
size $M\times M$ over the FS. Figure~\ref{MxM} demonstrates the dependence of
the ROC curves calculated for testing samples on~$M$ (for fixed $N=48$). It is
clearly seen that the effectiveness of training increases rapidly with a
decrease of~$M$, so that AUC reaches 0.997 for $M=8$.

As far as we can tell, two factors contributed to the dramatic increase
in the effectiveness of the CNN. Due to a comparatively small size of
the training set, it is easier for the network to recognize a short
signal on a small image segment than on a full image. Besides this, the
suggested way of organizing the input data strongly increases the number
of examples containing meteor signals in the training set (approximately
4 times for $M=8$ in comparison with $M=48$).

\section{Conclusions}

A simple convolutional network demonstrates high efficiency in recognition of
signals from meteors in the data of Mini-EUSO orbital fluorescence telescope
providing the input data is arranged in an appropriate way.  Accuracy of
classification is at least not worse than that obtained with conventional
algorithms. However, the CNN-based approach greatly reduces the time needed for
classification of newly arriving data.  Last but not least, applying the CNN
trained on data obtained with the time step of 40.96~ms to data recorded with
the time resolution 2.5~$\mu$s brought interesting results that will be reported
elsewhere. We believe that analysis of fluorescence telescope data with the help
of neural networks is very promising and will achieve further development in the
near future.  The experience of developing the discussed CNN will be applied to
the search for EAS-like events in the Mini-EUSO data and for EASs in the data of
the EUSO-SPB2 stratospheric experiment, the launch of which is scheduled for
2023~\cite{Eser:2021H6}.

\bigskip

The authors thank all members of the Mini-EUSO experiment for numerous helpful
discussions.  The study was supported by grant number 22-22-00367 of the Russian
Science Foundation (\url{https://rscf.ru/project/22-22-00367/}).

\bibliographystyle{model1a-num-names}
\bibliography{meteor_cnn}
\end{document}